# The Case for a Structured Approach to Managing Unstructured Data

AnHai Doan, Jeffrey F. Naughton,
Akanksha Baid, Xiaoyong Chai, Fei Chen, Ting Chen, Eric Chu, Pedro DeRose,
Byron Gao, Chaitanya Gokhale, Jiansheng Huang, Warren Shen, Ba-Quy Vuong

University of Wisconsin-Madison

## ABSTRACT

The challenge of managing unstructured data represents perhaps the largest data management opportunity for our community since managing relational data. And yet we are risking letting this opportunity go by, ceding the playing field to other players, ranging from communities such as AI, KDD, IR, Web, and Semantic Web, to industrial players such as Google, Yahoo, and Microsoft. In this essay we explore what we can do to improve upon this situation. Drawing on the lessons learned while managing relational data, we outline a structured approach to managing unstructured data. We conclude by discussing the potential implications of this approach to managing other kinds of non-relational data, and to the identify of our field.

## 1. MOTIVATION

Data management, broadly construed to encompass all kinds of data, has exploded in the past ten or so years. Once the province of large corporations, now virtually everyone with access to a computer deals with some form of online data; furthermore, even within large corporations, many more people deal with data and the data they deal with has more variety. A particularly prominent kind of data is *unstructured data*, which we take to include text documents, Web pages, emails, and so forth. In view of this, it is disconcerting that our community plays only a peripheral role in most of this data.

Of course, our community has long lamented that large chunks of the data space, especially those dealing with Web data, remain outside of our purview. But somehow today the problem is even more galling, perhaps because of the tremendous success of companies like Google, Yahoo, Microsoft, and myriad startups. These companies are making enormous amounts of money with the basic functionality of serving up data in response to user queries. This sounds like something we should care about and participate in. *The purpose of this essay is to speculate on how we might play a much more central role in the management of this kind of data in the future.* We believe that this presents an enormous opportunity for our community — perhaps the largest since our community started working on relational database management systems.

A perhaps somewhat surprising aspect of our proposal is that we are not really proposing a move away from structured data. Quite the contrary — we believe that our community's primary strength and contribution will remain in the direction of structured data. However, we are proposing a radical change in the source of the structured data. Rather than being created as structured data, we argue that in the future a main source of structured data should be unstructured data. That is, the structure we manage should be the structure that is currently hidden within unstructured data. As we will argue in the rest of the essay, dealing with this kind of structured data may require fundamental changes to the entire end-to-end systems we use to manage the data.

We also argue that if we are to be successful, *our data management model should be designed to allow human intervention at key points of the end-to-end data management process.* One way to put this is that we are not proposing that the data management community should solve an AI-complete problem. In particular, we do not mean to imply that our systems should automatically "understand" the meaning of unstructured documents. Rather, they should extract enough structure from these documents that humans can make deeper use of their content than they can with current IR-like systems. Humans may need to be involved in the loop at various points throughout the entire process, from extracting the structured data, to building the queries, and even to refining the entire process if the results they obtain are not what they wanted.

Or, as a reviewer of the first draft of this essay put it, we believe that human intervention is a fundamental piece of end-to-end systems to manage unstructured data. Consequently, if our community is to study such end-to-end systems (something that we should do and are well-equipped to do so), we would need to change what we know, and acquire and extend expertise traditionally left in the HCI community to tackle this fundamental piece, one that cannot be truly factored out and studied separately.

While we think the technical approach has merit, merely working on techniques to extract structure from unstructured documents and allowing for human interaction to help with the AI-complete problems encountered along the way will not be enough for success. *Retrospectively looking back on some key components in the success of relational systems may provide some insight as to what else is needed.* We can use this insight to both direct our efforts when we notice that some component is missing, and/or to decide that we might not be headed in the right direction if the creation of the missing component is out of our control.

The main components we have in mind are a data generation and exploitation model, an end-to-end system blueprint, and a business target. In particular, we will argue that to manage unstructured data effectively, we should develop a clear model of how the data is generated and exploited, and develop an end-to-end data management system blueprint that embodies the above model. This system blueprint can help rally the community and unify the disparate works, and hopefully enable rapid progress. Finally, we argue that for ultimate success, there needs to be an accompanying business community that ensures a cycle of "ideas to realistic prototype to commercial transfer back to ideas" for us, and speculate what that might look like.

The rest of this essay is structured as follows. Section 2 proposes that to maximize our impact, we should focus on generating and exploiting structure from unstructured data. Sections 3-5 then argue for the need of a data generation and exploitation model, an end-to-end system blueprint, and a business target, and speculate on these components. Section 6 discusses how what we propose here may be generalized to other types of data, and Section 7 concludes.

## 2. A FOCUS ON STRUCTURE

In managing unstructured data, if we stay at the *text* level and try to improve upon keyword search without changing the basic underlying approach, then we fear there is relatively little we can do.

Instead, we believe that our ambition should go beyond just better keyword search. To illustrate, consider Wikipedia today. With keyword search we cannot ask and obtain answers to questions such as "find the average March-September temperature in Madison, Wisconsin", even though the monthly temperatures appear on the Madison page. The fundamental reason is that to answer this question, the system must be able to *locate* the desired monthly temperatures, then *compute* their average, capabilities that are beyond today search engines. On the other hand, if we generate structure, such as ("month = September", "temperature = 70") from such data, then we can formulate and answer the above query over Wikipedia.

Consequently, we advocate that to maximize the benefits for users, we should focus on uncovering and exploiting the structure "hidden" in unstructured data.

This focus on structure will be much "in sync" with the broader research and industry landscape. Many communities, such as AI, Web, Semantic Web, IR, and KDD, have worked for years on extracting and exploiting structure from unstructured data, and they have recently been accelerating their efforts (e.g., see the WikiAI-08 workshop homepage at http://lit.csci.unt.edu/ wikiai08/index.php/Main_Page). In the industry, all major Web companies today are carrying out initiatives on extracting structure from unstructured data. The structure can then be exploited in a wide variety of applications, ranging from Web search, local search, portals, question answering forums, blog analysis and monitoring, user intelligence, marketing, to ad matching. More startups have also appeared recently in this area. Powerset, for example, is extracting and exploiting facts for question answering over Wikipedia, while Freebase is trying to extract then integrate all major publicly available data sets (e.g., Wikipedia, IMDB, US census data).

Our community however is uniquely well-equipped to enter this crowded arena, because the focus on structure plays to our traditional strength. We are the "Structure King", after all. As we will show in Sections 3-4, the structure focus raises many practical and interesting research problems. We are well suited to address them, by building on techniques that we have developed in the relational world. But we will have to examine and adapt them to deal with the new context (such as incorporating human intervention and managing uncertainty).

## 3. THE NEED FOR A DATA GENERATION AND EXPLOITATION MODEL

We now argue that to manage unstructured data effectively, a clear *data generation and exploitation model* (or *DGE model* for short) will have to emerge. Unfortunately, no such model has been identified by our community. We then speculate on such a model and explain its possible benefits. Section 4 then discusses the kind of data management systems we can build that embody such a model.

### 3.1 DGE Models

A DGE model explains the interaction between the data, the system, and the users. It explains how the data is generated inside the system, who the users are, what their information needs are, how they express the needs, and how they interact with the system to satisfy these needs.

For example, the DGE model we have (implicitly) used for relational data is as follows. To generate data, a user defines a schema, populates it with conforming data, and perhaps modifies the data by update transactions. To exploit the so-created data, a user poses a SQL query to the system, which produces an answer (the immediate "user" is often a program, but the model still holds). As another example, in the most popular DGE model for IR, data exploitation means a user's posing a keyword query to an IR system over a collection of text documents (given in the data generation step), then obtaining as answer a ranked list of the documents.

To manage any kind of data effectively, we argue that it is important to identify a good DGE model, one that captures most data management scenarios of interest. We can then build on the model to develop data models and management principles, as well as systems that embody such data models and principles. Furthermore, by capturing the fundamental interactions between the users, system, and data, such a model can help predict future trends. This in turn can help us identify problems that may be 5-10 years ahead of industry, thus putting us in a position to lead instead of reacting (as we further elaborate in Section 3.3).

### 3.2 Toward an DGE Model for Unstructured Data

Given the focus on structured data extracted from unstructured documents, the DGE models for relational data, keyword search, as well as those that have been proposed for the DB+IR context [1], are not appropriate. One main reason for this is that these models lack the incorporation of extraction activities. We now discuss what a reasonable DGE model for unstructured data might contain.

**Users:** We first consider the types of users that this model should handle. In the relational context, the DGE model in essence handles only sophisticated, SQL-knowing developers. Ordinary users (e.g., those who do not know SQL) play a very limited role. They interact with the database (to generate and query the data) simply by invoking canned SQL commands and queries (written by some developers) via relatively simple form interfaces.

In contrast, many applications involving unstructured data want to engage ordinary users actively in both the data generation and exploitation steps, a desire certainly heightened by the emergence of Web 2.0. For instance, an application involving Wikipedia may want ordinary users to participate in creating the wiki pages, as well as to be able to ask questions such as "find the average temperature of Madison" mentioned earlier. Consequently, a reasonable DGE model for unstructured data should allow not just sophisticated developers, but also ordinary users to participate in both the data generation and exploitation steps.

**Data Generation:** We have proposed to generate new data by *extracting* structured data from unstructured data, where in its simplest form this structured data is attribute-value pairs, such as temperatures, city names, locations, person names from Wikipedia.

Due to the nature of unstructured data, the extracted structured data will often be semantically heterogeneous. For example, the two different names "David Smith" and "D. Smith" extracted from Wikipedia may in fact refer to the same person, or attributes **location** and **address** extracted from two Wikipedia infoboxes may in fact match. Consequently, we will often have to perform an *information integration* step to resolve the semantic heterogeneity and unify the extracted structured data.

But automatic IE and II (i.e., information extraction and integration, respectively) often will not be 100% accurate. The fundamental reason is that they make many decisions based on the *data semantics*, and such semantics is often not adequately captured in the text, or adequately captured, but cannot be understood by the techniques (indeed, this is one of the key lessons learned from the IE and II work of the past two decades).

Given the above, applications often want to have a human in the loop, to help improve the accuracy of the underlying automatic IE/II techniques, as well as the accuracy of the final result. In the case of Wikipedia, for example, such a human user can correct semantic matches, or provide domain knowledge that helps improve matching accuracy. Consequently, our DGE model should allow the option of such human intervention (henceforth called *HI* for short).

Since we want ordinary users to be able to participate actively in the data generation process, it follows that we should allow not just developers, but also ordinary users in the HI step. Furthermore, the success of many Web 2.0 applications suggests that it may be highly beneficial to allow *a multitude of users*, instead of just a single one, to be able to provide feedback, in a *mass collaboration* fashion. Hence, it would be highly desirable for our DGE model to allow for this option.

Finally, many applications may want to generate structured data *incrementally, in a best-effort fashion*, as the user deems necessary (instead of generating all of them in one shot). For instance, a user looking for a new job may start out extracting only monthly temperatures from Wikipedia, as he or she only wants to do an average temperature comparison across U.S. cities. Later if the user wants to examine only cities with at least 500,000 people, then he or she may want to also extract city populations, and so on. Consequently, our DGE model should allow the structured data to be generated in an incremental, best-effort fashion, should the application choose to do so.

**Data Exploitation:** We turn now to the data exploitation step. Recall that we want both sophisticated and ordinary users to be able to exploit the derived structured data. Consider again the question $Q =$ "find the average temperature of Madison" in the Wikipedia example. Suppose we have extracted the monthly temperatures, then a sophisticated user can immediate formulate $Q$ as a structured query (e.g., in SQL), and obtain an answer from the system.

An ordinary user however does not know SQL and most likely would just want to start with a keyword query, such as "average temperature Madison". In this case it would be highly desirable for the system to guide the user somehow to a structured-query reformulation of $Q$. One way to do so is to "guess" and show the user several structured queries using, say, form interfaces, then ask the user to select the appropriate one.

In general, then, our DGE model should allow users to start in whatever data-exploitation mode they deem comfortable (e.g., keyword search, structured querying, browsing, visualization), then help them move seamlessly into the mode that is ultimately appropriate for their information need. Furthermore, users often start with an ill-defined information need, then refine it during the exploration process. Our model should effortlessly support this as well.

**Summary:** We have argued that a good DGE model for unstructured data should use a combination of IE, II, and HI to generate structured data from the originally unstructured data, in a potentially mass collaboration, best-effort fashion. The model should allow a broad range of data exploitation modes (e.g., keyword search, structured querying, browsing, visualization, monitoring), as well as seamless transition from one mode to another, in an iterative fashion through interaction with the user.

### 3.3 Benefits of the Proposed DGE Model

Once we have developed a DGE model for unstructured data, such as described above, we can benefit from it in two important ways. First, we can build on it to develop data models and management principles that are appropriate for the unstructured data context.

For instance, we have run into examples of what we think could be interesting data management principles that involve HI. The idea is that in many cases we have run into situations where it is very easy for users to recognize something that fits their needs, yet very difficult for them to generate this something without help. For example, in II, often narrowing the set of potential matches to a manageable number allows users to spot the correct match, when they would be swamped by the total number of potential matches and would not succeed if they had no automated assistance. Similarly, it appears that users are much better at recognizing when a query form matches their information need than at writing the equivalent SQL query from scratch. We think this is just one aspect of a fundamental principle that may even be related to the underlying issues in P vs. NP (ease of discovery of a solution vs. ease of verification of its correctness.)

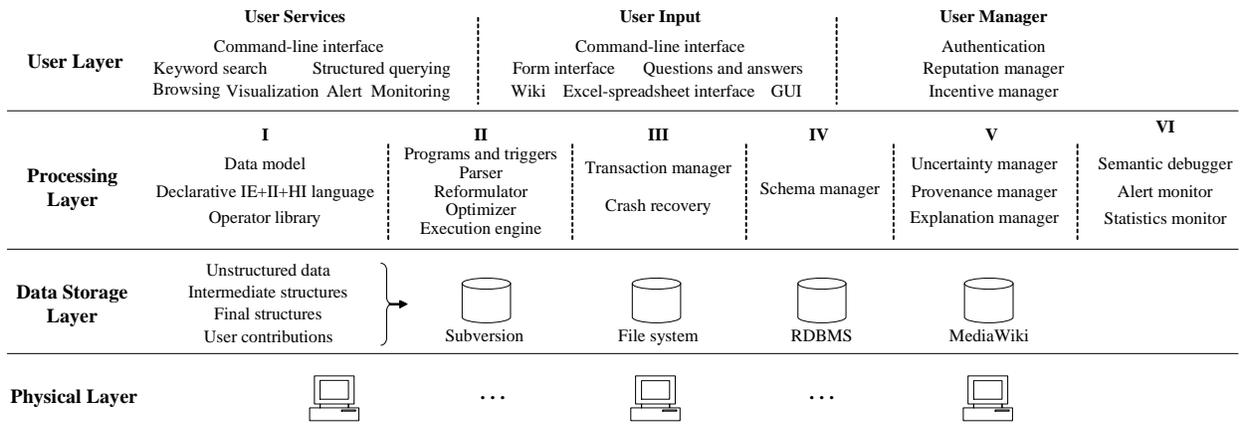

Figure 1: A possible architecture for a general system to manage unstructured data.

As another example, we have found that there are tasks that would be very difficult for automatic techniques, and yet easy for human users. Examples include recognize if a particular person is present in a picture, and if a form interface is a gateway to an online store (as opposed to, say, being a subscription interface). Using this principle, during the data generation step, we can try to isolate and expose such tasks to HI to maximize their accuracy.

Another potentially important benefit we can derive from the DGE model is to use it to predict future trends. To illustrate, the vast majority of academic and industrial work on unstructured data has so far focused only on *extracting* structured data. Our proposed DGE model, however, suggests that if such work continues, sooner or later they would run into a particular *exploitation* problem, namely, how to enable ordinary users to easily ask structured queries over the derived structured data. Attacking such problems can then help put us in a position to lead, instead of reacting to current events.

## 4. THE NEED FOR AN END-TO-END SYSTEM BLUEPRINT

Having discussed desirable properties for an DGE model for unstructured data, we now turn to the issue of building systems that embody such a model.

We start by noting that, in retrospect, the relational world received a huge benefit from the early creation of complete prototype systems such as System R and Ingres. With these systems as examples and context, an entire community arose working on improving their performance and broadening their scope. This unified a lot of what would otherwise be disparate work, helped guide research, enabled rapid progress, and resulted in real-world systems that magnified the dissemination of the products of our community's efforts.

In the unstructured data world, we argue that it is highly desirable to have a similar example system, one that can rally the community and unify the work, and hopefully enable rapid progress. In fact, given the many CS communities playing today in the data management arena, we should perhaps focus on the system building angle as a distinguishing aspect: our community builds *end-to-end scalable data management systems*. We do not have such a systems today. But we can speculate on what such a system should contain, given the above DGE model.

In what follows we discuss such a possible system, as depicted in Figure 1. This system consists of four layers: physical layer, data storage layer, processing layer, and user layer.

**The Physical Layer:** This layer contains hardware that runs the data generation and exploitation steps. Given that IE and II are often very computation intensive and that many applications involve a large amount of unstructured data, we need parallel processing in the physical layer. A popular way to achieve this is to use a computer cluster (as shown in the figure) running Map-Reduce-like processes.

**The Data Storage Layer:** This layer stores all forms of data: the original unstructured data, intermediate structured data derived from it (kept around for example for debugging, HI, or optimization purposes), the final structured data, and user contributions. These different forms of data have very different characteristics, and may best be kept in different storage devices, as depicted in the figure (of course, other choices are possible, such as developing a single unifying storage device).

For example, if the unstructured data is retrieved daily from a collection of Web sites, then the daily snapshots will overlap a lot, and hence may be best stored in a device such as Subversion, which only stores the "diff" across the snapshots, to save space. As another example, the system often executes only sequential reads and writes over intermediate structured data, in which case such data can best be kept in the file systems. As yet another example, if the system allows concurrent editing by multiple users on the final structure, then this structure may be best stored in an RDBMS, to ensure fast and correct concurrency control.

**The Processing Layer:** This layer is responsible for specifying and executing the data generation processes. At the heart of this layer is a data model, a declarative language (over this data model) that combines IE, II, and HI, and a library of basic operators (see Part I of this layer in the figure).

Developers can then use the language and operators to write declarative IE+II+HI programs that specifies how to extract, integrate, and curate the data. These programs can be parsed, reformulated (to subprograms that are executable over the storage devices in the data storage layer), optimized, then executed (see Part II in the figure). Note that developers may have to write domain-specific operators, but the framework makes it easy to use such operators in the programs.

The remaining four parts, Parts III-VI in the figure, contain modules that provide support for the data generation process. Part III handles transaction management and crash recovery. Part IV manages the schema of the derived structure. Since this structure often is generated in an incremental, best-effort fashion (see Section 3.2), in many cases the schema will evolve over time. Hence, Part IV will likely have to deal with schema evolution challenges.

Part V handles the uncertainty that arise during the IE, II, and HI processes. It also provides the provenance and explanation for the derived structured data.

Part VI contains an interesting module called the semantic debugger. This module learns as much as possible about the application semantics. It then monitors the data generation process, and alerts the developer if the semantics of the resulting structure is not "in sync" with the application semantics. For example, if this module has learned that the monthly temperature of a city cannot exceed 130 degrees, then it can flag an extracted temperature of 135 as suspicious. This part also contain modules to monitor the status of the entire system and alert the system manager if something appears to be wrong.

**The User Layer:** This layer allows users (ordinary and sophisticated alike) to exploit the data as well as provide feedback into the system. The part "User Services" contains all common data exploitation modes, such as command-line interface (for sophisticated users), keyword search, structured querying, etc. The part "User Input" contains all common interfaces that can be used to solicit user feedback, such as command-line interface, form interface, wiki, etc. (see the figure).

We note that modules from both parts will often be combined, so that the user can also conveniently provide feedback while querying the data, and vice versa. Finally, this layer also contains modules that authenticates users, manage incentive schemes for soliciting user feedback, and manage user reputation (e.g., for mass collaboration).

As described, we believe such a system should be sufficiently general to be applicable to many real-world applications, ranging from personal information management, community information management, scientific data management, local search, Web search, to online ad management. It should also encompass many existing IR, IE, and II systems, and can be viewed as a next logical step in extending current DB+IR system efforts [1].

It should also be clear from the description that developing such a system raises numerous challenges, such as IE, II, HI, large-scale data processing, efficient storage of text data, declarative query languages, optimization, schema evolution, uncertainty management, provenance, translating keyword queries into structured ones, and so on.

As such, such a system blueprint can potentially serve as a unifying point for many current research challenges (as well as a starting point for novel ones). To address these challenges, we can build on techniques that we have developed in the relational world, but we will have to examine and adapt them to the new contexts (e.g., handling HI and text data).

## 5. THE NEED FOR A BUSINESS TARGET

Developing the technical approach – as we have proposed – is all well and good. But merely working on models and systems will not be enough for success. We believe that a robust data management community cannot be built in a vacuum without any associated target business use of the data. For one reason, the community will need the financial support that only comes with a compelling business application. For another reason, students will be unlikely to train to work in such a community if there are no jobs for them when they finish. But even for non-financial reasons we need a business target, so that we can create the virtuous cycle of ideas to prototypes to commercial distribution back to ideas. The existence of a successful relational database management industry has played an essential role in the success of our community to date, and we think an equivalent industry will be essential going forward.

This is not to say that the research community should function as developers for the business side of the community. The relationship between the research community and the business community may vary over time, sometimes the two will be close, other times they will diverge for awhile before reconnecting. But without such a connected business community the research community will not reach its potential.

Currently, there is no such business community based upon managing unstructured data by extracting the hidden structure. This raises the question of what we as researchers should do about this. For most of us it is not within our expertise to decipher what such a business community should look like, nor is it within our ability to force one to arise. But this doesn't mean that the presence of absence of such a business community is irrelevant to our work.

Perhaps an approach that makes sense is for us to propose strawman models for what a business might look like. Undoubtedly we will get the details wrong, but such a model might still prove valuable as a source of guidance for our efforts. Also, if we can't even envision a business around the kinds of systems we are proposing, then it is likely that while we may have found interesting research projects, the systems are unlikely to provide the thrust for a new expansion of the size and relevance of our data management community.

What might this industry look like? We think that our best bet is to focus on managing Web data, since there are well-proven business models there. Once we have developed good systems, we can try other domains (just like RDBMSs were first developed for enterprises, but are now used in many other domains).

What can we do on the Web? The most well-known application of managing unstructured data is Web search, carried out by large Web companies. It is difficult to build a realistic Web search prototype, simply because due to the complexity of Web search, no open source system is close to what the companies have built, and also because the Web is simply too large for most research groups to manage. Furthermore, Web companies will understandably not give out their code nor provide access to all of their enormous computational resources. So while we can potentially make impact here (e.g., by studying how structured data can help Web search), it may be limited and work well only for a small number of researchers. If the future is just more Web search, we may have only limited opportunity to be relevant.

We argue, however, that the future is not likely to look like the present. Web 2.0 has demonstrated that it is possible to develop many small-to-medium-size applications, put them out there, then attract users that use them to manage data. Examples include Wikipedia, Del.icio.us, Flickr, YouTube, and numerous social search engines (e.g., Wikia Search), among many others.

Capitalizing on this trend, Web companies large and small have found a new business model: they develop such applications (and often also the hosting platform), then invite developers to use them to build compelling Web services that attract eyeballs, then split the ad revenue with the developers. An example is Yahoo! Search BOSS (Build Your Own Search Service) platform, which developers, start-ups, and large Internet companies can use to build and launch Web-scale search products that utilize the entire Yahoo! Search index. Another example is Google Knol platform, which anyone can use to host a group to edit wikis, then split the ad revenue with Google.

This trend of "we will help your develop and deploy Web applications, then in return share revenue with us" appears likely to continue. If so, it provides a possible ecosystem within which our envisioned new "structure from unstructured data, with humans in the loop" industry could grow. We can develop applications that would make it very easy for developers or ordinary users to extract and exploit structured data over some slice of the Web. These applications can then be plugged into such a hosting platform, for real-world testing. The applications can address a broad range of problems, such as managing personal data, building portals, wikis, intranets, and so on. Since they handle only a slice, not the whole Web, we envision that most research groups can build and manage them (especially if such applications can be made open source, so that we can build on each other's efforts, instead of starting from the scratch).

The above scenario offers an interesting vision for the evolution of the Web: the Web will become increasingly structured, but in a bottom up fashion. This will happen because there will be increasingly more applications that try to help users to generate structured data and exploit the fruits. Our community could be at the center of this new, increasingly structured Web, as we help develop such applications.

## 6. BEYOND UNSTRUCTURED DATA

So far we have made a case for a structured approach to managing unstructured data, such as emails, text, Web pages. We believe, however, that this approach may work for other kinds of data as well, with suitable modifications. One example is image data, from which we want to extract and then manipulate real-world objects (e.g., table, car, person). Another example is sensor data from which we want to infer real-world events (e.g., someone has entered the room). Yet another example is heterogeneous data, i.e., data that come from a collection of disparate sources; here we may want to infer semantic matches among the data elements, then use the matches to integrate the data into a coherent whole.

In all of these cases, we want to extract some kind of higher-level structure from the underlying raw data. Such extracted structured data will often be semantically heterogeneous, suggesting the need for integration techniques. The inherent imperfection of extraction and integration in turn suggests that it may be desirable to have humans in the loop, and so on. The end system then may end up looking quite similar to the kind of systems we have discussed for unstructured data, and hence can potentially benefit from work in that area.

## 7. CONCLUDING REMARKS

Unstructured data is big and we are risking letting the opportunities to manage it go by. In this essay we have argued for a structured approach to manage such data, and have outlined the components and the challenges of the approach. We regard this approach as a baseline. Our hope is that this essay will spark further discussions on how to improve this baseline into an effective approach to managing unstructured data for our entire community.

Beyond unstructured data, throughout the essay we have also alluded to questions regarding the identify of our field. These questions have been perennial at our community gatherings. But with the entrance of new fields (e.g., AI, Web, Semantic Web, KDD, IR) into the data management arena, and the rapid rise of large Web players (e.g., Google, Yahoo, Microsoft), answering such questions has become more urgent. This essay has provided a possible answer, namely, we can count among our unique characteristics a focus on *structure* and on *building end-to-end scalable data management system*. We hope that the essay can spark further discussions on this matter as well.

**Acknowledgment:** This work is supported by NSF grants SCI-0515491, Career IIS-0347943, an Alfred Sloan fellowship, an IBM Faculty Award, and a grant from Microsoft. We thank the reviewers for invaluable comments on an earlier draft of this essay.